\documentclass[notitlepage,12pt]{iopart}

\usepackage{graphics}
\usepackage{graphicx}
\usepackage{epsf}
\usepackage[dvips]{color}
\usepackage{pifont}
\input{colordvi.sty}

\begin{document}
\definecolor{blue}{rgb}{0.15,0.1,0.7}
\definecolor{red}{rgb}{0.7,0.1,0.15}
\definecolor{green}{rgb}{0.15,0.7,0.15}
\newcommand{\blue}[1]{{\textcolor{blue}{ \small #1}}}
\newcommand{\red}[1]{{\textcolor{red}{ \small #1}}}

\title[]{Readout of a single electron spin in a double quantum dot using a quantum point contact}

\author{Jian-Ping Zhang$^1$, Shi-Hua Ouyang$^{1, 2}$, Chi-Hang Lam$^2$, J Q You$^1$}

\address{$^1$Department of Physics
and Surface Physics Laboratory (National Key Laboratory), Fudan
University, Shanghai 200433, China}
\address{$^2$Department of Applied Physics, Hong Kong Polytechnic
University, Hung Hom, Hong Kong, China}
\date{\today}
\begin{abstract}
We study the dynamics of a single electron spin in a double quantum
dot (DQD) and its readout via a quantum point contact (QPC). We
model the system microscopically and derive rate equations  for the
reduced electron density matrix of the DQD. Two cases with one and
two electrons in the DQD are studied. In the one-electron case, with
different Zeeman splittings in the two dots, the electron spin
states are distinctly characterized by a constant and an oscillatory
current through the QPC. In the two-electron case, the readout of
the spin state of the electron in one of the dots called the qubit
dot is essentially similar after considering hyperfine interactions
between the electrons and the nuclear spins of the host materials
and a uniform magnetic field applied to the DQD. Moreover, to ensure
that an electron is properly injected into the qubit dot, we propose
to determine the success of the electron injection from the
variations of the QPC current after applying an oscillating magnetic
field to the qubit dot.

\end{abstract}
%
\section{Introduction}

A single or a small number of electron spins 
confined in a semiconductor quantum dot (QD) has become a subject of
considerable interest, partly motivated by potential applications in
quantum information processing. Because a electron spin in a QD can
have a relatively long decoherence time, it is a promising candidate
for realizing a qubit \cite{Hanson2003,DLosspra57}, the basic unit
of a quantum computer.
For a single electron spin, the dephasing time is found in theory to
be $\rm{\mu s}$ in both GaAs \cite{Yao06,Yao07} and InAs dots
\cite{Liu07}. Experimental results show that the ensemble dephasing
time of an electron spin has the order of $ns$ \cite{Bracker05,
Koppens05}. Moreover, the spin relaxation time in a large GaAs dot
is found to be about $1-10~\mu s$ at a moderately low temperature of
$10$K \cite{Cheng06}.
Indeed, recent experiments have demonstrated that spins in QDs can
be used to carry quantum information
\cite{TFujisawa,JMElzerman11,ACjohnson}. For both applications in
quantum computing and fundamental research, the readout of qubit
states based on electron spin is a centrally important issue
\cite{M.A. Nielsen}. However, due to the weak magnetic moment
associated with the electron spin, it is difficult to directly
measure the electron spin states. A possible solution is to
correlate the spin states to charge states, and the measurement of
the charge on the dot will provide information about the original
spin states \cite{DLosspra57}. This can be implemented using a
quantum point contact (QPC), which is a charge detector and can be
used to determine the number variation of the electrons confined in
the QD.

Recently, the  readout of electron spin states in a QD has been
realized using such a spin-charge conversion
\cite{JMElzerman11,RHanson2005}. For the experiment in
Ref.~\cite{JMElzerman11}, a QD is connected to an electron
reservoir. Applying an external magnetic field, gate voltages are
applied so that the electron confined in the dot can tunnel to the
reservoir if its spin is down. (A spin-up electron cannot tunnel in
this case).  A nearby QPC is used to detect the electron number
variation in this QD and can determine the electron spin state in
the QD. Also, Engel et al. \cite{Hans-AndreasE93} proposed various
implementations of the readout process based on a double quantum dot
(DQD). Barrett and Stace \cite{SDBarrett96} proposed a electron spin
readout approach using a microwave field and an inhomogeneous Zeeman
splitting across the DQD.

In the present paper, we study two implementations for reading-out
electron spin states based on a DQD coupled to a QPC.  Also, we
explain the effects of static and oscillating magnetic fields on the
electron spin states. The first implementation involves a single
electron in the DQD. The readout of the spin states is based on the
difference Zeeman splittings in the two QDs. In the second
implementation, two electrons are allowed in the DQD. The Pauli
exclusion principle and hyperfine interactions between the electrons
in the DQD and the nuclear spins in the host materials enable the
readout of the electron spin states. These are interesting examples
for implementing readout of the electron spin states.
A potential advantage of our proposal is that the readout
manipulation can easily be switched on (off) by decreasing
(increasing) the tunneling barrier between the two dots through
varying the gate voltages. Thus, the readout process can be
implemented only when needed. This is important in quantum
information processing.
To understand the underlying physics from a microscopic point of
view, we derive a set of rate equations describing the electron
dynamics of the DQD system. Based on these rate equations, we
calculate the QPC current and illustrate that the QPC current
behaves differently for different spin states.

The paper is organized as follows. In Sec.~2, we model the system
when only one electron is confined in the DQD. A set of Bloch-type
rate equations are derived to describe the detailed measurement
processes for the electron spin states in the qubit dot. In Sec.~3,
we study the measurement of the electron spin states in the qubit
dot when two electrons are confined in the DQD. Sec.~4 is the
conclusion.

\section{Readout of single electron spin: one electron in DQD}
\subsection{Theoretical model}
\begin{figure}
\epsfxsize 10.50cm \centerline{\epsffile{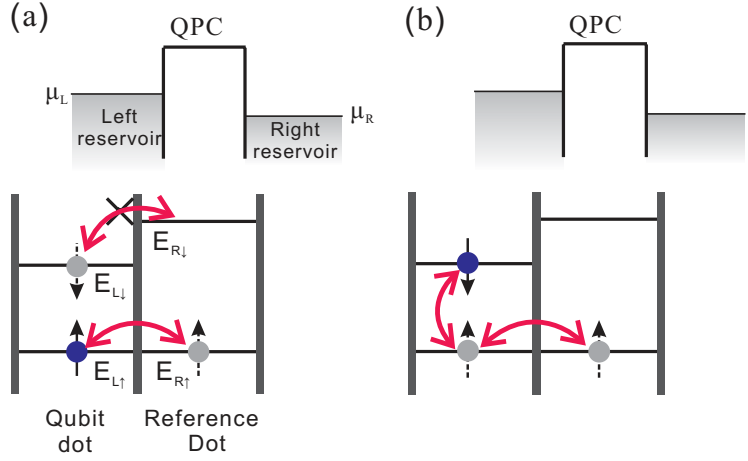}}
\caption{(Color online). Schematic diagram of a double quantum dot
(DQD) and a quantum point contact (QPC) with only one electron in
the DQD. The left (qubit) dot of the DQD is coupled to a right
(reference) dot via hopping. The nearby QPC is used as a detector
measuring the number variation of electron in the reference dot. An
energy-level detuning of the electron spin states is generated using
two external magnetic fields in the two dots. (a) Gate voltages are
adjusted to keep $E_{L\uparrow}=E_{R\uparrow}$, so that the hopping
of the spin-up electron between the left and right dot is allowed.
Moreover, the energy-level detuning for the spin-down electron is
much larger than the hoping strength, i.e.,
$E_{R\downarrow}-E_{L\downarrow}\gg\Omega_0$, and hence the hopping
of the spin-down electron between the two dots is forbidden. (b) The
hopping blockade for the spin-down electron is lifted by applying a
transverse magnetic field $B(t)=B_L^x\cos(\omega_c t)$ to the left
dot, which flips the electron spin.}\label{fig1}
\end{figure}

We first discuss a scheme to detect the electron spin states in the
case with only one electron confined in the DQD. As schematically
shown in Figure.~\ref{fig1}, the whole system consists of a DQD and
a QPC. The left dot is used as a qubit dot, in which the electron
spin is expected to be readout. The right dot is used as a reference
dot. The QPC is capacitively coupled to the right dot and serves as
a readout device. The electron number variation in the right dot
induces a change in the barrier in the QPC. This leads to a
variation of the current through the QPC, which can be used to
indicate the occupation of the right dot \cite{Mfield70}.

The Hamiltonian of the whole system is given by
\begin{equation}
H=H_{\rm{DQD}}+H_{\rm{QPC}}+H_{\rm{int}}+H_{\rm{rf}},
\end{equation}
with
\begin{eqnarray}
H_{\rm{DQD}}\!&\!=\!&\!\sum_{i\sigma}{E_{i\sigma}c_{i\sigma}^{\dag}c_{i\sigma}}%
+\sum_{\sigma}{\Omega_{0}(c_{L\sigma}^{\dag}c_{R\sigma}+c_{R\sigma}^{\dag}c_{L\sigma}}),
\nonumber\\
&&~~~~~~~~~~~~~~~~~~~~~~~~~~~~~~~~~~~~\label{H-DQD-one}
\end{eqnarray}
where $i=L,R$ denote the left and right dots, and $c_{i\sigma}^\dag$
($c_{i\sigma}$) is the creation (annihilation) operator of electron
with spin $\sigma$ in the $i$th QD. $\Omega_{0}$ denotes the hopping
amplitude between the two dots and here it is assumed to be
spin-independent. We have denoted the energy levels in the $i$-th
dot by $E_{i\uparrow(\downarrow)}=E_i\mp\frac{1}{2}\Delta_{i}^z$,
with $\Delta_{i}^z=g\mu_B B_i^z$, where $E_i$ is the orbital energy
level of the QD and $B_i^z$ is an externally applied magnetic field
in the $i$-th dot along the $z$ direction. Here, $g$ is the
effective gyromagnetic factor and $\mu_B$ is the Bohr magneton. We
have chosen the unit $\hbar=1$. The Hamiltonian of the QPC reads
\begin{equation}
H_{\rm{QPC}}=\sum_{\alpha k}{E_{\alpha k}a_{\alpha
k}^{\dag}a_{\alpha k}}
+\sum_{lrk}\Omega_{lr}{\left(a_{lk}^{\dag}a_{rk}+a_{rk}^{\dag}a_{lk}\right)},\label{H-QPC}
\end{equation}
where $a_{\alpha k}^\dag$ ($a_{\alpha k}$) is the creation
(annihilation) operator of an electron with momentum $k$ in
reservoir $\alpha$ ($\alpha=l,r$). $H_{\rm{int}}$ gives the
electrostatic interaction between the DQD and the QPC:
\begin{equation}
H_{\rm{int}}=\sum_{lrk \sigma}
\delta{\Omega_{lr}}c_{R\sigma}^{\dag}c_{R\sigma}(a_{lk}^{\dag}a_{rk}+a_{rk}^{\dag}a_{lk}).\label{H-int}
\end{equation}
An electron spin resonance (ESR) magnetic field is applied in the
$x$ direction at the left dot, leading to a term
\begin{eqnarray}
H_{\rm{rf}}=\Delta_x(t)(c_{L\uparrow}^+c_{L\downarrow}+\rm{H.c.})
\end{eqnarray}
with $\Delta_x(t)=\frac{1}{2}g\mu_B B_L^x\cos(\omega_c t)$. This ESR
magnetic field generates spin flipping when it is resonant with the
Zeeman splitting on the left dot, i.e., $\omega_c=g\mu_B B_L^z$.

The spin-up and spin-down states $|\!\uparrow_L\rangle$ and
$|\!\downarrow_L\rangle$ in the left dot constitute the basis states
of a qubit. The right dot works as a reference dot, the electron
occupation of which is measured by the nearby QPC. Energy detuning
for the spin-up (down) electron is
$\varepsilon_{\uparrow(\downarrow)}=E_{R\uparrow(\downarrow)}-E_{L\uparrow(\downarrow)}$.
We assume that the Zeeman splittings $\Delta_i^{z}$ are different in
the two dots.
This can be realized, e.g., by applying a micro-size permanent
magnet near one dot of the DQD \cite{MPLadriere}.
This leads to a difference
$\varepsilon_{\uparrow}-\varepsilon_\downarrow=\Delta_{R}^z-\Delta_{L}^z$
in the energy level splittings for the spin-up and spin-down
electrons. In our consideration, gate voltages are adjusted to keep
$\varepsilon_\uparrow\approx0$, so that a spin-up electron can hop
back and forth between the two dots. Furthermore, we also assume
$\varepsilon_\downarrow\gg\varepsilon_\uparrow,\Omega_0$, so that
hopping is forbidden for spin-down electron. However, this spin
blockade can be lifted by an ESR magnetic field. Here, in the
one-electron case, the effects of the nuclear magnetic fields in the
two dots are neglected. This is because the Zeeman splitting in each
dot is much larger than the nuclear field in the x and y directions.
Moreover, the z component of the nuclear field only shifts the
energy level and can be included in the Zeeman splitting.

The physical picture of the electron spin readout is as follows. We
first inject an electron with either up or down spin into the qubit
(left) dot. An initially spin-up electron in the qubit dot can hop
into the reference dot. This will lead to a variation of the current
through the QPC. In contrast, for an initially spin-down electron in
the qubit dot, it will remain stationary because
$\varepsilon_\downarrow\gg\Omega_0$. As a result, no variation in
the QPC current occurs. Therefore, one can determine the initial
electron spin state based on the variation of the current through
the QPC. However, in practical experiments, the injection of
electrons into the DQD may not be always successful. Without any
electron in the DQD, there is also no variation in the QPC current.
Thus, this simple implementation cannot distinguish between the
cases with zero or one spin-down electron. To solve this problem, as
will be shown below, one can apply an ESR magnetic field in the
qubit dot. The ESR magnetic field induces spin flipping in the left
dot. If there is a spin-down electron, it can be converted to the
spin-up state by the ESR field and then hop onto the right dot.
Therefore, a current variation will be observed in the QPC. In
contrast, the QPC current will remain unchanged in the zero-electron
case even in the presence of the ESR field.

\subsection{Bloch-type rate equation}

To describe the physical processes quantitatively, we derive a set
of Bloch-type rate equations for the reduced density matrix
$\sigma(t)$ of the DQD system. Following Gurvitz et
al.\cite{Gurvitz1,Gurvitz2}, we write the wave function of the whole
system in the occupation representation as
\begin{eqnarray}\label{1b}
|\Psi{(t)}\big\rangle&=&
\sum_\sigma\Big[b_{L\sigma}(t)c_{L\sigma}^{\dag}+b_{R\sigma}(t)c_{R\sigma}^{\dag}
\nonumber\\
&&+\sum_{lr}b_{L\sigma lr}(t)c_{L\sigma}^{\dag}a^{\dag}_ra_l
+\sum_{lr}b_{R\sigma lr}(t)c^{\dag}_{R\sigma}a^{\dag}_ra_l
\nonumber\\
&&+\sum_{l<l^\prime,r<r^\prime}b_{L\sigma
l{l^\prime}r{r^\prime}}(t)c_{L\sigma}^{\dag}
a_r^{\dag}a_{r^\prime}^{\dag}a_la_{l^\prime}
\nonumber\\
&&+\sum_{l<l^\prime,r<r^\prime}b_{R\sigma l{l^\prime}r{r^\prime}}(t)
c_{R\sigma}^{\dag}a_r^{\dag}a_{r^\prime}^{\dag}a_la_{l^\prime}+\ldots
\Big]|0\rangle, \nonumber\\
&&~~~~~~~~~~~~~~~~~~~~~~~~~~~
\end{eqnarray}
where $b_j(t)$, $j=L\sigma, R\sigma, L\sigma lr, R\sigma lr, ...$
are the time-dependent probability amplitudes to find the system in
the corresponding states. For example, $b_{L\sigma lr}(t)$ denotes
the probability amplitude for the state with an electron having
tunnelled through the QPC barrier (from the left reservoir to the
right one) at time $t$, and an extra electron with spin $\sigma$
staying in the left dot. The vacuum state $|0\rangle$ corresponds to
the state where there is no extra electron in the DQD and all the
energy levels up to the Fermi energies $\mu_{L}$ and $\mu_R$ of the
two reservoirs of the QPC are occupied by electrons.

The relevant electron states of the DQD span a four-dimensional
Hilbert space. We adopt the notations
$|1\rangle\equiv|\!\!\uparrow_L\rangle$
and $|2\rangle\equiv|\!\downarrow_L\rangle$
for the left-dot states, as well as
$|3\rangle\equiv|\!\uparrow_R\rangle$
and $|4\rangle\equiv|\!\downarrow_R\rangle$ for the right-dot
states. A diagonal element $\sigma_{ii}(t)$ ($i=1,2,3,4$) of the
reduced matrix represents the occupation probability of the state
$|i\rangle$, while an off-diagonal element $\sigma_{ij}(t)$
characterizes the quantum coherence.
Each $\sigma_{ii}$ is
further given by
\begin{equation}
\sigma_{ii}(t)=\sigma_{ii}^{(0)}+\sigma
_{ii}^{(1)}+\sigma_{ii}^{(2)}+\ldots,
\end{equation}
where $\sigma_{ii}^{(n)}$ is the probability that the DQD is at
state $|i\rangle$ after $n$ electrons have tunnelled from the left
reservoir of the QPC to the right one. In this notation, we have,
for example
\begin{eqnarray}
\sigma_{11}^{(0)}&=&|b_{L\uparrow}(t)|^2,~\sigma_{11}^{(1)}=\sum_{lr}|b_{L\uparrow
lr}(t)|^2, ~\sigma_{11}^{(2)}=\sum_{l<l',r<r'}|b_{L\uparrow
ll'rr'}|^2,~\ldots.
\end{eqnarray}
%
The current flowing through the QPC is
\begin{equation}
I_{\rm{QPC}}(t)=e\frac{dN(t)}{dt},
\end{equation}
where $N(t)$ is the number of electrons transported to the right
reservoir of the QPC at time $t$. Accordingly, we have
\begin{eqnarray}
I_{\rm{QPC}}(t)=\sum_{n,i}n\dot{\sigma}_{ii}^{(n)}(t).
\end{eqnarray}

Substituting the many-body wave function of the whole system into
the Schr\"{o}dinger equation
$i|\dot{\psi}(t)\rangle=H|\psi(t)\rangle$, one gets a set of
differential equations for the probability amplitudes $b_j(t)$. In
the nonequilibrium transport in the QPC with a large voltage bias,
following Refs.~\cite{Gurvitz2} and \cite{Ouyang}, the Bloch-type
rate equations for the reduced density matrix $\sigma(t)$ of the DQD
are derived by integrating the degrees of freedom of the QPC
reservoirs. By summing $\dot{\sigma}^{(n)}(t)$ over $n$, the rate
equations for the diagonal elements are given by
\begin{eqnarray}\label{OneElectron-diagonal}
\dot{\sigma}_{11}(t)&=&i\Omega_{0}(\sigma_{13}-\sigma_{31})
+i\Delta_x(t)(\sigma_{12}-\sigma_{21}),\nonumber\\
\dot{\sigma}_{22}(t)&=&i\Omega_{0}(\sigma_{24}-\sigma_{42})
+i\Delta_x(t)(\sigma_{21}-\sigma_{12}),\nonumber\\
\dot{\sigma}_{33}(t)&=&i\Omega_{0}(\sigma_{31}-\sigma_{13}),\nonumber\\
\dot{\sigma}_{44}(t)&=&i\Omega_{0}(\sigma_{42}-\sigma_{24}),
\end{eqnarray}

The rate equations for the off-diagonal elements are
\begin{eqnarray}
\dot{\sigma}_{12}(t)&=&i(E_{L\downarrow}-E_{L\uparrow})\sigma_{12}
-i\Omega_{0}\sigma_{32}+i\Omega_{0}\sigma_{14}+i\Delta_x(t)(\sigma_{11}-\sigma_{22})
,
\nonumber\\
\dot{\sigma}_{13}(t)&=&i(E_{R\uparrow}-E_{L\uparrow})\sigma_{13}
+i\Omega_{0}(\sigma_{11}-\sigma_{33})-i\Delta_x(t)\sigma_{23}\nonumber\\
&&-\frac{\Gamma_d}{2}\sigma_{13}
-\frac{\chi}{2}\sigma_{11}-\frac{\chi}{2}\sigma_{33},
\nonumber\\
\dot{\sigma}_{14}(t)&=&i(E_{R\downarrow}-E_{L\uparrow})\sigma_{14}
+i\Omega_{0}\sigma_{12}-i\Omega_{0}\sigma_{34}-i\Delta_x(t)\sigma_{24}\nonumber\\
&&-\frac{\Gamma_d}{2}\sigma_{14}
-\frac{\chi}{2}\sigma_{12}-\frac{\chi}{2}\sigma_{34},
\nonumber\\
\dot{\sigma}_{23}(t)&=&i(E_{R\uparrow}-E_{L\downarrow})\sigma_{23}
+i\Omega_{0}\sigma_{21}-i\Omega_{0}\sigma_{43}
-i\Delta_x(t)\sigma_{13}\nonumber\\
&&-\frac{\Gamma_d}{2}\sigma_{23}
-\frac{\chi}{2}\sigma_{21}-\frac{\chi}{2}\sigma_{43},
\nonumber\\
\dot{\sigma}_{24}(t)&=&i(E_{R\downarrow}-E_{L\downarrow})\sigma_{24}
+i\Omega_{0}(\sigma_{22}-\sigma_{44})
-i\Delta_x(t)\sigma_{14}\nonumber\\
&&-\frac{\Gamma_d}{2}\sigma_{24}
-\frac{\chi}{2}\sigma_{22}-\frac{\chi}{2}\sigma_{44},
\nonumber\\
\dot{\sigma}_{34}(t)&=&i(E_{R\downarrow}-E_{R\uparrow})\sigma_{34}+i\Omega_{0}\sigma_{32}
-i\Omega_0\sigma_{14}, \nonumber\\
&&~~~~~~~~~~~~~~~~~~~~~~~~~~~\label{OneElectron-off}
\end{eqnarray}
where $\Gamma_d=\left(\sqrt{D'}-\sqrt{D}\right)^2$ is the dephasing
rate induced by the QPC detector \cite{Gurvitz2}.
Here we have defined
\begin{eqnarray}
D=2\pi\rho_L\rho_R\Omega^{2}V_d, D'=2\pi\rho_L\rho_R\Omega'^{2}V_d,
\end{eqnarray}
and
\begin{equation}
\chi=\frac{\Lambda}{V_d}\left(\frac{\Omega}{\Omega'}+\frac{\Omega'}{\Omega}-2\right),
\end{equation}
with $\Lambda=2\pi\rho_L\rho_R\Omega'\Omega_{0}\Omega{V_d}.$ In
Eq.~(\ref{OneElectron-off}), the terms proportional to $\chi$ are
due to  the inclusion of higher-order terms of
$O(\Omega^2\Omega_0/V_d^2)$ \cite{Ouyang}. Also, we assume that the
tunneling couplings depend weakly on the energy, so that
$\Omega_{lr}(E_l,E_r)\equiv\Omega$, and
$\Omega_{lr}+\delta\Omega_{lr}\equiv\Omega'$. $V_d=\mu_L-\mu_R$ is
the voltage bias applied on the QPC and $\rho_L(\rho_R)$ is the
density of states in the left (right) reservoir of the QPC. From
Eq.~(\ref{OneElectron-off}), one can see that $\Gamma_d$
characterizes the exponential damping of the off-diagonal density
matrix elements. Now, the QPC current is given by
\begin{equation}\label{c8}
I_{\rm{QPC}}(t)=I_0[\sigma_{11}(t)+\sigma_{22}(t)]+I_1[\sigma_{33}(t)+\sigma_{44}(t)],
\end{equation}
where $I_0\equiv D$ is the current flowing through the QPC when the
right dot of the DQD is empty, while $I_1\equiv D'$ is the QPC
current when the right dot is occupied by one electron. Since
$I_0\neq I_1$ in general, one can determine the electron occupation
of the right dot from the variation of the QPC current.
\begin{figure}
\epsfxsize 12cm \centerline{\epsffile{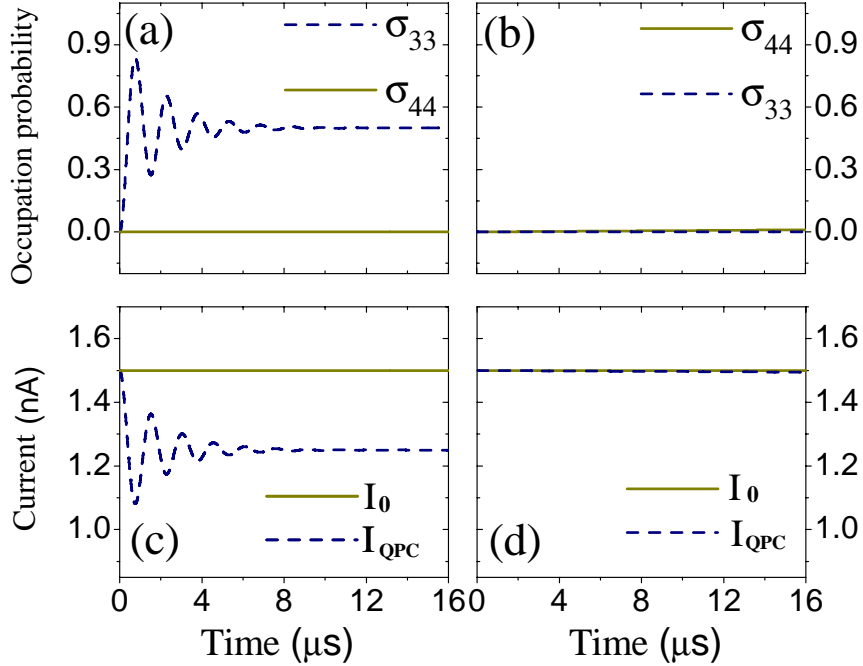}}
\caption{(Color online). Time evolution of the electron occupation
probability in the right dot for (a) spin-up $|\uparrow_L\rangle$
and (b) spin-down $|\downarrow_L\rangle$ electron states in the left
dot. (c) and (d) Time evolution of the QPC currents corresponding to
(a) and (b), respectively. We have set the parameters as
$\Omega_0=0.25~\mu {\rm eV}$, $\chi=0.0025~\mu {\rm eV}$ and
$\Gamma_d=60$~M\rm {Hz}.}\label{fig2}
\end{figure}

We first consider the case without an ESR oscillating magnetic
field, i.e., $\Delta_x(t)=0$. Using
Eqs.~(\ref{OneElectron-diagonal}) and (\ref{OneElectron-off}), one
can numerically calculate the occupation probabilities
$\sigma_{ii}$, $i=1$ to $4$. A typical value for the hopping
coupling between the two dots in experiments is
$\Omega_0=0.25~\mu$ev\cite{Wiel03}. We have taken parameters so that
the initial current of the QPC is $I_0=1.5$~nA if the right dot is
empty \cite{JMElzerman11}, while it equals $I_1=1$~nA if there is an
electron in the right dot. First, consider the case that a spin-up
electron is injected into the left dot. Figure~\ref{fig2}(a) shows
the calculated occupation probability of the electron in the right
dot. The corresponding current flowing through the QPC is given in
Figure.~{\ref{fig2}}(c). It shows that the current $I_{\rm{QPC}}$
starts from the initial value $I_0$, and then decreases, oscillates
and finally converges to a value other than $I_0$. In addition,
oscillations in both the occupation probability and the QPC current
are observed. These results from the fact that the spin-up electron
can tunnel back and forth between the dots. In contrast, if the
electron injected into the left dot is spin-down, it cannot hop into
the right dot because $\varepsilon_{\downarrow}\gg\Omega_0$. The
electron occupation probability in the right dot is hence zero [see
Figure.~\ref{fig2}(b)]. The down spin is also reflected in
Figure.~\ref{fig2}(d), where the QPC current remains unchanged.
Accordingly, one can distinguish between the two initial electron
spin states from the variation of the QPC current. In short, if the
QPC current decreases from its initial value, the initial spin state
is spin up. Alternatively, if the initial spin state is spin-down in
the left dot, the QPC current remains unchanged.

\begin{figure}
\epsfxsize 12cm \centerline{\epsffile{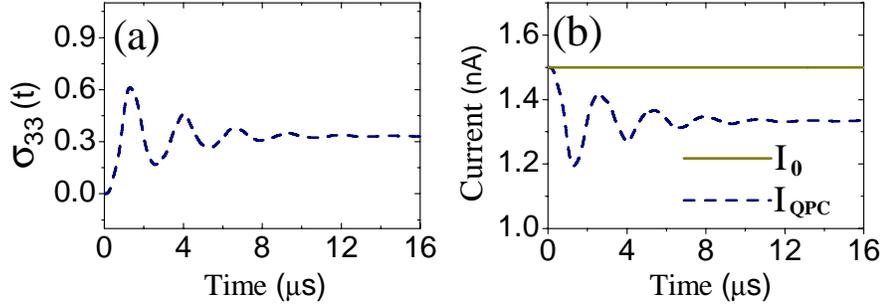}}
\caption{(Color online). (a)Time evolution of the occupation
probability in the spin-up state $|\uparrow_R\rangle$ in the right
dot in the presence of an oscillating magnetic field. Initially, the
electron in the left dot is in the spin-down state
$|\downarrow_L\rangle$. (b)The corresponding time evolution of the
QPC current. We have taken $\Delta_x=0.3$~$\mu\textrm{eV}$, and
$\omega=0.5$~$\mu\textrm{eV}$.}\label{fig3}
\end{figure}
We have repeated the calculation by considering an additional ESR
magnetic field. Without such a field, we cannot distinguish between
the case with no electron in the left dot from that with a spin-down
electron as discussed in Sec.~2.1. Both give rise to no QPC current
variation. In the presence of an ESR field on the left dot (i.e.,
$\Delta_x\neq0$), for the zero-electron case, the QPC current
remains unchanged. However, for an initial spin-down electron in the
left dot $|\!\downarrow_L\rangle$, it can flip to the spin-up state
$|\!\uparrow_L\rangle$, induced by the ESR oscillating magnetic
field. In contrast, the spin-up electron can hop into the right dot
[see Figures.~\ref{fig3}(a)]. This leads to a variation in the QPC
current [see Figures.~\ref{fig3}(b)]. Therefore, these two cases can
now be distinguished.

\section{Readout of single electron spin: two electrons in DQD}
\subsection{Theoretical model}

\begin{figure}
\epsfxsize 10.50cm \centerline{\epsffile{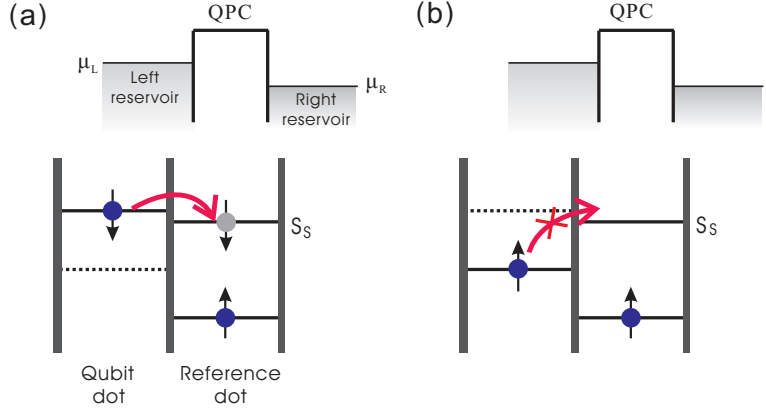}}
\caption{(Color online). Schematic diagram of a double quantum dot
(DQD) and a quantum point contact (QPC) with two electrons in the
DQD. One spin-up electron is initially kept in reference dot by
properly adjusting the gate voltages. (a) A spin-down electron in
the left dot can always hop into the right dot after taking
hyperfine interactions into account. (b) Transport of a spin-up
electron is forbidden due to Pauli exclusion. 
}\label{fig4}
\end{figure}

We now study the readout of the electron spin states in the left
(qubit) dot assuming that an additional electron initially occupies
the right (reference) dot (see Figure.~\ref{fig4}), as in a recent
experiment \cite{FHLK766}. We further assume that the gate voltages
of the dots are tuned so that no two electrons can simultaneously
stay in the qubit dot. Thus, the two relevant occupation
configurations correspond to two electrons in the right dot or one
electron in each dot. The total Hamiltonian is
\begin{equation}
H\!=\!H_{\rm{QPC}}+H_{\rm{DQD}}+H_{\rm{int}},\label{H-two}
\end{equation}
where $H_{\rm{QPC}}$ is the Hamiltonian of the QPC in the
two-electron case, which has the same form as ~Eq.~(\ref{H-QPC}),
but with $\Omega_{lr}$ replaced by $\Omega'_{lr}$.
The Hamiltonian of the isolated DQD system after considering both
inter- and intradot Coulomb interactions now becomes
\begin{equation}
H_{\rm{DQD}}\!=\!H_0+H_{\rm{spin}},
\end{equation}
where
\begin{eqnarray}
H_{0}\! &=&\!\sum_{i\sigma }E_{i}c_{i\sigma }^{+}c_{i\sigma }
+\Omega_0\sum_{\sigma }(c_{L\sigma }^{+}c_{R\sigma }+\rm{H.c.})
\nonumber\\
&+&\sum_{i}U_{i}n_{i\uparrow }n_{i\downarrow}
+U_{LR}\sum_{\sigma\sigma'}n_{L\sigma}n_{R\sigma'}.\label{eq4}
\end{eqnarray}

In the absence of a net nuclear polarization, randomly oriented and
fluctuating nuclear spins in the host materials give rise to
effective magnetic fields $\mathbf{B_{NL}}$ and
$\it{\mathbf{B_{NR}}}$ in the left and right dot, respectively. They
results from different local environments for the electrons in the
respective dots. However, nuclear fields change with a nuclear spin
relaxation time scale of the order $1$~s, which is much longer than
any time scales characterizing the transport processes of electron.
These nuclear effective fields can thus be regarded as static fields
in our discussion \cite{WAC70,RdeS,DP15}. Therefore, we can describe
the influence of the magnetic fields on the electron spins in the
DQD by
\begin{eqnarray}
H_{\rm{spin}}\!&\!=\!&\!
g\mu_B\mathbf{B}_{\rm{NL}}\cdot{\mathbf{S}}_{\rm{L}}
+g\mu_B\mathbf{B}_{\rm{NR}}\cdot{\mathbf{S}}_{\rm{R}}
\nonumber \\
&&+g\mu_BB_{\rm{ext}}^{\rm{z}}(S_{\rm{L}}^{\rm{z}}+S_{\rm{R}}^{\rm{z}})
+g\mu_BB_{\rm{L}}^{\rm{x}}\cos(\omega_c{t})S_{\rm{L}}^{\rm{x}},\nonumber\\
&&~~~~~~~~~~~~~~~~~~~~~~~~~\label{eq5}
\end{eqnarray}
where $\mathbf{S}_{\rm{L}}$ and $\mathbf{S}_{\rm{R}}$ correspond to
the electron spin in the left and right dots, respectively. The
third term in Eq.~(\ref{eq5}) is the Zeeman splitting caused by an
external perpendicular field. The last term is an ESR oscillating
magnetic field in the $x$ direction. In the present discussion, we
assume that the ESR oscillating magnetic field is only applied on
the qubit dot.

The relevant electronic states for the DQD span a five dimensional
Hilbert space. The basis set consists of double-dot triplets
$|1\rangle \equiv |T_{+}\rangle \!=\!|\!\uparrow _{L}\uparrow
_{R}\rangle$,
$|2\rangle \equiv |T_{-}\rangle \!=\!|\!\downarrow _{L}\downarrow
_{R}\rangle$,
and $|3\rangle \equiv
|T_{0}\rangle\!=\!\frac{1}{\sqrt{2}}(|\!\uparrow _{L}\downarrow
_{R}\rangle +|\!\downarrow _{L}\uparrow _{R}\rangle)$,
double-dot singlet $|4\rangle \equiv |S_D\rangle\!
=\!\frac{1}{\sqrt{2}}(|\!\uparrow _{L}\downarrow _{R}\rangle
-|\!\downarrow _{L}\uparrow _{R}\rangle)$,
and single-dot singlet $|5\rangle \equiv |S_S\rangle\!
=\!\frac{1}{\sqrt{2}}(|\!\uparrow _{R}\downarrow _{R}\rangle
-|\!\downarrow _{R}\uparrow _{R}\rangle)$.
Single-dot triplet states are excluded due to their much higher
orbital energies \cite{Ashoori93,ACjohnson}.  In this
representation,
$H_{\rm{DQD}}$ is rewritten as
\begin{eqnarray}
&&H_{\rm{DQD}}=\sum_{i=1,2,3,4,5}E_i|i\rangle\langle i|
\nonumber\\
&&+\frac{g\mu_B}{\sqrt{2}}\big[(B_s^x+iB_s^y)|3\rangle\langle1|
+(B_s^x-iB_s^y)|3\rangle\langle2|+\rm{H.c.}\big]
\nonumber\\
&&+\frac{g\mu_B}{\sqrt{2}}\big[(-B_d^x-iB_d^y)|4\rangle\langle1|
+(B_d^x-iB_d^y)|4\rangle\langle2|+\rm{H.c.}\big]\nonumber\\
&&+\Omega_0(|4\rangle\langle5|+|5\rangle\langle4|)
+g\mu_BB_d^{z}\big(|3\rangle\langle4|+|4\rangle\langle3|\big)
\nonumber\\
&&+\Omega _{1}\cos (\omega _{c}t)\big[|3\rangle \langle1|+|3\rangle
\langle2|-|4\rangle\langle1|+|4\rangle \langle2|+\rm{H.c.}\big],
\nonumber\\
&&\label{H-DQD-two}
\end{eqnarray}
where
\begin{math}
\mathbf{B}_d\!=\!\frac{1}{2}(\mathbf{B}_{\rm{NL}}-\mathbf{B}_{\rm{NR}}),
~\mathbf{B}_s\!=\!\frac{1}{2}(\mathbf{B}_{\rm{NL}}+\mathbf{B}_{\rm{NR}})+
{B}^z_{{ext}}\tilde{\mathbf{z}},
\end{math}
and $ \Omega_{1}=\frac{1}{2\sqrt{2}}g\mu_BB_{L}^{x}.$ We have also
introduced energy levels given by
\begin{eqnarray}
&&E_{1,2}\!=\!E_3\mp g\mu_BB_s^z,~E_{3,4}\!=\!E_{L}+E_R+U_{LR},
\end{eqnarray}
and
\begin{equation}
E_{5}\!=\!2E_R+U_R.
\end{equation}
A critical step in the readout is the hopping to the right dot,
where there is a non-zero Coulomb energy barrier
\begin{eqnarray}
\label{Delta} \Delta = E_5 - E_4 =U_R-U_{LR} - (E_L-E_R),
\end{eqnarray}
for the second electron at the right dot if the intra-dot repulsion
$U_R$ dominates.

The interaction Hamiltonian between the DQD and the QPC is
\begin{equation}\label{H-DQD-int}
H_{\rm{int}}=\sum_{lrk\sigma}\delta\Omega'_{lr}c^{\dag}_{{R}\sigma}
c_{{R}{\sigma}}c^{\dag}_{{R}\bar{\sigma}}c_{{R}{\bar{\sigma}}}(a_{lk}^{\dag}a_{lk}+a_{rk}^{\dag}a_{rk}),
\end{equation}
In the singlet-triplet representation, it can be written as
\begin{equation}\label{H-int-two}
H_{\rm{int}}=\sum_{lrk}\delta\Omega'
|5\rangle\langle5|(a_{lk}^{\dag}a_{rk}+a_{rk}^{\dag}a_{lk}).
\end{equation}
Similar to the one-electron case, the hopping amplitude
$\Omega'_{lr}$ of the QPC and its change $\delta\Omega'_{lr}$ by
either adding or removing an electron in the right dot are assumed
to be energy-independent, so that
$\Omega'_{lr}(E_l,E_r)\equiv\Omega'$ and
$\Omega'_{lr}+\delta\Omega'_{lr}\equiv\Omega''$.
%
As discussed in the one-electron case, the left QD is a qubit dot
and the right dot is a reference dot. The nearby QPC works as a
detector to measure the number change of electrons in the reference
dot. When there is one electron staying in the reference dot, the
current flowing through the detector is $I_1$. For double occupancy
in the reference dot, the detector current becomes $I_2$, where
$I_2<I_1$ because of the increased QPC barrier induced by the
additional electron. As a result, the variation of the electron
number in the reference dot can be reliably detected from the
current change of the detector.

We first briefly discuss the readout processes of the qubit states.
We assume that the electron that is always kept in the right dot is
spin-up. This can be realized by injecting an unpolarized electron
and wait until a time interval much longer than the typical
relaxation time of a single electron spin. It will relax to its
ground spin-up state due to its coupling to the outside
environments. An additional electron is then injected into the qubit
dot and its spin state will be readout. For a spin-down qubit
electron, the initial total spin in the $z$ direction is $S_z=0$.
The DQD system takes either the double electron state $|3\rangle$ or
$|4\rangle$ with an equal probability of $1/2$. If the state taken
is $|4\rangle$, the electron can directly hop onto the right dot.
This hopping is described by the $\Omega_0$ term in
Eq.~(\ref{H-DQD-two}). Otherwise, if it is $|3\rangle$, the electron
can transit to state $|4\rangle$, as allowed by the $B_d^z$ term in
Eq.~(\ref{H-DQD-two}). Then hopping into the right dot becomes
possible similarly.  Therefore, the qubit electron can always hop
onto the reference dot on the right, leading to the state
$|5\rangle$ for $S_z=0$. Due to this hopping, the QPC current
changes from $I_1$ to another value and this indicates the spin-down
qubit state.

In contrast, for a spin-up qubit electron leading to a total spin
component $S_z=1$, it form the triplet state $|1\rangle$ with the
spin-up electron in the reference dot. Because of the application of
a large external magnetic field $B_{\rm{ext}}^z\gg\sqrt{\langle
B_N^2\rangle}$, this triplet state $|1\rangle$ is far away in energy
from other states. Thus, it is decoupled to states $|3\rangle$ or
$|4\rangle$ and the electron in the left dot cannot hop to the right
dot. Therefore, the current flowing through the QPC remains constant
at $I_1$ and this indicates that the initial qubit state is spin-up.

\subsection{Bloch-type rate equation}

To reveal the quantum dynamics of electron states in the DQD system,
we derive a set of Bloch-type rate equations for the reduced density
matrix $\sigma(t)$ of the DQD, also using the technique developed by
Gurvitz et al\cite{Gurvitz1}. We assume the high Zeeman splitting
limit, i.e., $B_{\rm{ext}}^z \gg\sqrt{\langle B_N^2\rangle}$, in
order to suppress the effect of the nuclear fields. Spin flips
caused by hyperfine interactions are then negligible. The many-body
wave function $|\Psi(t)\rangle$ of the whole system in the
singlet-triplet basis is given by
\begin{eqnarray}\label{wavefunction-two}
|\Psi(t)\big\rangle\!&\!=\!&\!\sum_{i=1,2,3,4,5}\big[b_i(t)c_i^+
%
%
+\sum_{lr}b_{ilr}(t)c_{ilr}^+a_r^{\dag}a_l
\nonumber\\
&&+\sum_{l<l',r<r'}b_{ill'rr'}(t)c_{i}^+a_r^{\dag}a_{r'}a_la_{l'}
+\ldots\Big]|0\big\rangle. \nonumber\\
&&~~~~~~~~~~~~~~~~~~~~~~~~~~~~~~~~~~~~~~~
\end{eqnarray}
where $|0\rangle$ is the vacuum state and $b_j(t)$ is the
time-dependent probability amplitudes of the corresponding state
$|j\rangle$. For example, when $j\!=\!ilr$, with $i=1,2, ..$ or $5$,
$b_j(t)$ is the probability amplitude of the state with the DQD
system at state $|i\rangle$ while one electron has already passed
through the QPC at time $t$. In addition, we have used $c_i^\dag$
($c_i$), which denotes the creation (annihilation) operator for
state $|i\rangle$ in the DQD system.

Substituting the wave function $|\Psi(t)\rangle$
[Eq.~(\ref{wavefunction-two})] into the Schr\"{o}dinger equation
$i|\dot{\Psi}(t)\rangle\!=\!H|\Psi(t)\rangle$, and tracing over the
reservoir states of the QPC, we obtain a set of Bloch-type rate
equations for the reduced density matrix $\sigma(t)$ of the DQD
system:
\begin{eqnarray}\label{Diagonal-two}
\dot{\sigma}_{11}(t)&=&iD_-\sigma_{41}-iS_-\sigma_{31}
-iD_+\sigma_{14}+iS_+\sigma_{13},
\nonumber\\
\dot{\sigma}_{22}(t)&=&-iD_+\sigma_{42}-iS_+\sigma_{32}
+iD_-\sigma_{24} +iS_-\sigma_{23},
\nonumber\\
\dot{\sigma}_{33}(t)&=&i\Omega_d^z(\sigma_{34}-\sigma_{43})
+iS_+\sigma_{32}+iS_-\sigma_{31} -iS_-\sigma_{23}-iS_+\sigma_{13} ,
\nonumber\\
\dot{\sigma}_{44}(t)&=&i\Omega_d^z(\sigma_{43}-\sigma_{34})
+iD_+\sigma_{42} -iD_-\sigma_{41}
+i\Omega_0(\sigma_{45}-\sigma_{54})\nonumber\\
&&-iD_-\sigma_{24} +iD_+\sigma_{14},
\nonumber\\
\dot{\sigma}_{55}(t)&=&i\Omega_0(\sigma_{54}-\sigma_{45}),
\end{eqnarray}
and
\begin{eqnarray*}\label{TwoElectron-off}
\dot{\sigma}_{12}(t)&=&i(E_2-E_1)\sigma_{12}+iD_-\sigma_{14}
+iS_-\sigma_{13} +iD_-\sigma_{42}
-iS_-\sigma_{32}\nonumber,\\
\dot{\sigma}_{13}(t)&=&i(E_3-E_1)\sigma_{13}+iD_-\sigma_{43}
-iS_-\sigma_{33} +iS_+\sigma_{12}
\nonumber\\
&&+iS_-\sigma_{11}+i\Omega_d^z\sigma_{14}\nonumber,\\
\dot{\sigma}_{14}(t)&=&i(E_4-E_1)\sigma_{14}+iD_-(\sigma_{44}
-\sigma_{11})-iS_-\sigma_{34}+iD_+\sigma_{12}
\nonumber\\
&&+i\Omega_0\sigma_{15}+i\Omega_d^z\sigma_{13}\nonumber,\\
\dot{\sigma}_{15}(t)&=&i(E_5-E_1)\sigma_{15}+i\Omega_0\sigma_{14}
+iD_-\sigma_{45} -iS_-\sigma_{35}
-\frac{1}{2}\Gamma_d'\sigma_{15}\nonumber\\
&&-\frac{1}{2}\chi'\sigma_{14}.\nonumber\\
\dot{\sigma}_{23}(t)&=&i(E_3-E_2)\sigma_{23}-iD_+\sigma_{43}
-iS_+(\sigma_{33}-\sigma_{22})
\nonumber\\
&&+iS_-\sigma_{21}+i\Omega_d^z\sigma_{24}\nonumber,\\
\dot{\sigma}_{24}(t)&=&i(E_4-E_2)\sigma_{24}-iD_+(\sigma_{44}
-\sigma_{22})-iS_+\sigma_{34} -iD_-\sigma_{21}
\nonumber\\
&&+i\Omega_0\sigma_{25}+i\Omega_d^z\sigma_{23}\nonumber,\\
\dot{\sigma}_{25}(t)&=&i(E_5-E_2)\sigma_{25}+i\Omega_0\sigma_{24}
-iD_+\sigma_{45}-iS_+\sigma_{35}
-\frac{1}{2}\Gamma_d'\sigma_{25}\nonumber\\
&&-\frac{1}{2}\chi'\sigma_{24}\nonumber,\\
\dot{\sigma}_{34}(t)&=&-iS_-\sigma_{24} -iS_+\sigma_{14}
-iD_-\sigma_{31}+iD_+\sigma_{32}
-i\Omega_d^z(\sigma_{44}-\sigma_{33})+i\Omega_0\sigma_{35}
\nonumber,\\
\dot{\sigma}_{35}(t)&=&i(E_5-E_3)\sigma_{35}+i\Omega_0\sigma_{34}
-iS_-\sigma_{25}-iS_+\sigma_{15} -i\Omega_d^z\sigma_{45}
\nonumber\\
&&-\frac{1}{2}\Gamma_d'\sigma_{35}-\frac{1}{2}\chi'\sigma_{34}\nonumber,\\
\dot{\sigma}_{45}(t)&=&i(E_5-E_4)\sigma_{45}+i\Omega_0(\sigma_{44}
-\sigma_{55})-iD_-\sigma_{25}+iD_+\sigma_{15}
\nonumber\\
&&-i\Omega_d^z\sigma_{35}-\frac{1}{2}\Gamma_d'\sigma_{45}
-\frac{1}{2}{\chi}'(\sigma_{44}+\sigma_{55})\nonumber,\\
\end{eqnarray*}
Here the detector-induced dephasing rate is
 $\Gamma_d'=\left(\sqrt{D''}-\sqrt{D'}\right)^2$,
with
\begin{equation}
D''=2\pi\rho_L\rho_R{\Omega''}^{2}V_d,~~D'=2\pi\rho_L\rho_R{\Omega'}^{2}V_d.\nonumber
\end{equation}

Also, we have defined
\begin{eqnarray}
\chi'=\frac{\Lambda'}{V_d}(\frac{\Omega'}{\Omega''}+\frac{\Omega''}{\Omega'}-2),~~\Omega_d^z=g\mu_BB_d^z,\nonumber\\
D_\pm(t)=\Omega_d^\pm+\Omega_1(t),~~S_{\pm}(t)=\Omega_s^\pm+\Omega_1(t),
\end{eqnarray}
where
\begin{eqnarray}
\Lambda'=2\pi\rho_L\rho_R\Omega''\Omega_{0}\Omega'{V_d},~~\Omega_1(t)=\Omega_{1}\cos(\omega_ct),\nonumber\\
\Omega_d^{\pm}=\frac{g\mu_B}{\sqrt{2}}(B_d^x\pm
iB^y_d),~~\Omega_s^{\pm}=\frac{g\mu_B}{\sqrt{2}}(B_s^x\pm iB^y_s).
\end{eqnarray}
The QPC current is given by
\begin{equation}\label{Current-two}
I(t)=I_1[\sigma_{11}(t)+\sigma_{22}(t)+\sigma_{33}(t)+\sigma_{44}(t)]+I_2\sigma_{55}(t).
\end{equation}
where $I_1$ ($I_2$) is the stationary current through the QPC when
the right dot is occupied by one electron (two electrons).


\subsection{Results and analysis}

We have numerically integrated the rate equations and obtained the
time-dependent density matrix elements. As discussed in Sec.~3.1,
since there is always a spin-up electron in the reference dot, the
injection of a spin-up electron into the qubit dot forms a
double-dot triplet state $|1\rangle=|\!\uparrow_L\uparrow_R\rangle$
in the DQD system. In contrast, if the injected electron is
spin-down, the DQD system initially takes the state
$|\!\downarrow_L\uparrow_R\rangle$. Thus, after injecting an
electron into the left dot, the possible experimental initial states
of the DQD system are $|1\rangle$ and
$|\!\downarrow_L\uparrow_R\rangle$. In order to show how the current
through the QPC changes for different initial states of the DQD
system, we assume that the DQD system initially takes the state
$|1\rangle$ or $|\!\downarrow_L\uparrow_R\rangle$.
\begin{figure}
\epsfxsize 12cm \centerline{\epsffile{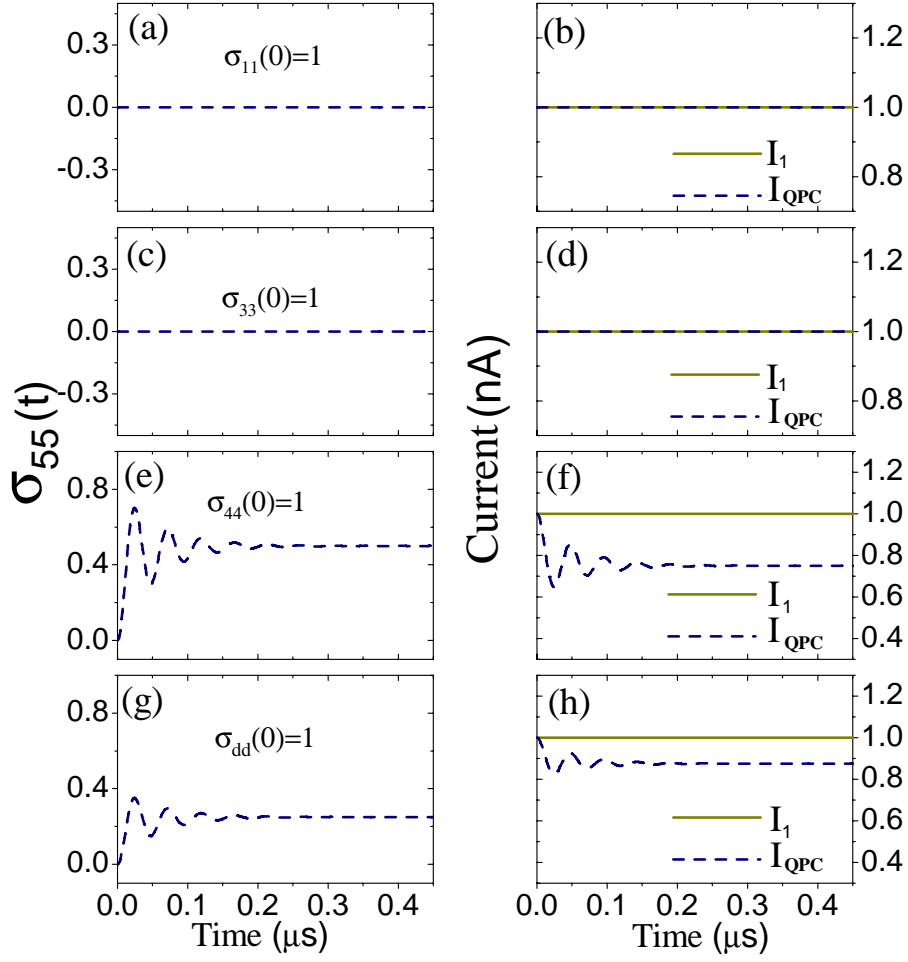}}
\caption{(Color online). Time evolution of the occupation
probability $\sigma_{55}(t)$ for the single-dot singlet state
$|S_S\rangle$ with four different initial conditions: (a)
$\sigma_{11}(0)=1$, (c) $\sigma_{33}(0)=1$, (e) $\sigma_{44}(0)=1$
and (g) $\sigma_{dd}(0)=1$, where the hyperfine interaction is not
considered. (b), (d), (f) and (h) The corresponding QPC currents.
Here we have chosen the following parameters: $B_{\rm{NL}}^{x, y,
z}=(0, 0, 0)$~m$\textbf{T}$, $B_{\rm{NR}}^{x,y,z}=(0, 0, 0)$
m$\textbf{T}$, $\omega_c=3.75$~$\mu\textrm{eV}$,
$\Delta=0.25$~$\mu\textrm{eV}$, $\Omega_0=0.25$~$\mu\textrm{eV}$,
$\chi'=0.0025$~$\mu\textrm{eV}$,
$\Omega_{1}=0.375$~$\mu\textrm{eV}$, and
$\Gamma_d'=60$~MHz.}\label{fig5}
\end{figure}

The initial state $|\!\downarrow_L\uparrow_R\rangle$ is a
superposition of the double-dot triplet state $|3\rangle$ and the
double-dot singlet state $|4\rangle$, i.e.,
\begin{equation}
|\!\downarrow_L\uparrow_R\rangle=\frac{1}{\sqrt{2}}(|3\rangle-|4\rangle).\label{superposition}
\end{equation}
Here the state $|4\rangle$ is coupled to the single-dot singlet
state $|5\rangle$ directly via hopping coupling, while the state
$|3\rangle$ is coupled to $|5\rangle$ via the intermediate state
$|4\rangle$ (where the transition from $|3\rangle$ to $|4\rangle$ is
induced by the $B_d^z$ term). To reveal the contributions by
different components, we also take the state $|3\rangle$ or
$|4\rangle$ as the initial state to study the time evolution of the
current through the QPC.

In our numerical calculations regarding the initial state
$|\!\downarrow_L\uparrow_R\rangle$, we rewrite the rate equations
(\ref{Diagonal-two}) and (\ref{TwoElectron-off}) in the occupation
representation defined by the basis states
$|a\rangle,\;|b\rangle,\;|c\rangle,\;|d\rangle$, and $|e\rangle$,
where
\begin{eqnarray}
&&|a\rangle\equiv|\!\uparrow_L\uparrow_R\rangle=|1\rangle,\;\;|b\rangle\equiv|\!\downarrow_L\downarrow_R\rangle=|2\rangle,\nonumber\\
&&|c\rangle\equiv|\!\uparrow_L\downarrow_R\rangle=\frac{1}{\sqrt{2}}(|3\rangle+|4\rangle),\;\;
|d\rangle\equiv|\!\downarrow_L\uparrow_R\rangle=\frac{1}{\sqrt{2}}(|3\rangle-|4\rangle),\nonumber\\
&&|e\rangle\equiv\frac{1}{\sqrt{2}}(|\!\uparrow_R\downarrow_R\rangle-|\!\downarrow_R\uparrow_R\rangle)=|5\rangle.
\end{eqnarray}
With these new basis states, one can express
$\sigma_{ij}\equiv\langle i|\sigma|j\rangle$ ($i,j=1$ to $5$) using
$\sigma_{\mu\nu}\equiv\langle\mu|\sigma|\nu\rangle$
($\mu,\nu=a,b,c,d$, and $e$), e.g.,
\begin{eqnarray}
&&\sigma_{13}=\frac{1}{\sqrt{2}}(\sigma_{ac}+\sigma_{ad}),\;\;\sigma_{23}=\frac{1}{\sqrt{2}}(\sigma_{bc}+\sigma_{bd}),\nonumber\\
&&\sigma_{33}=\frac{1}{2}(\sigma_{cc}+\sigma_{cd}+\sigma_{dc}+\sigma_{dd}),\;\;%
\sigma_{43}=\frac{1}{2}(\sigma_{cc}+\sigma_{cd}-\sigma_{dc}-\sigma_{dd}).
\end{eqnarray}
In this way, we can transform equations (\ref{Diagonal-two}) and
(\ref{TwoElectron-off}) into the rate equations in the occupation
representation.

\begin{figure}
\epsfxsize 12cm \centerline{\epsffile{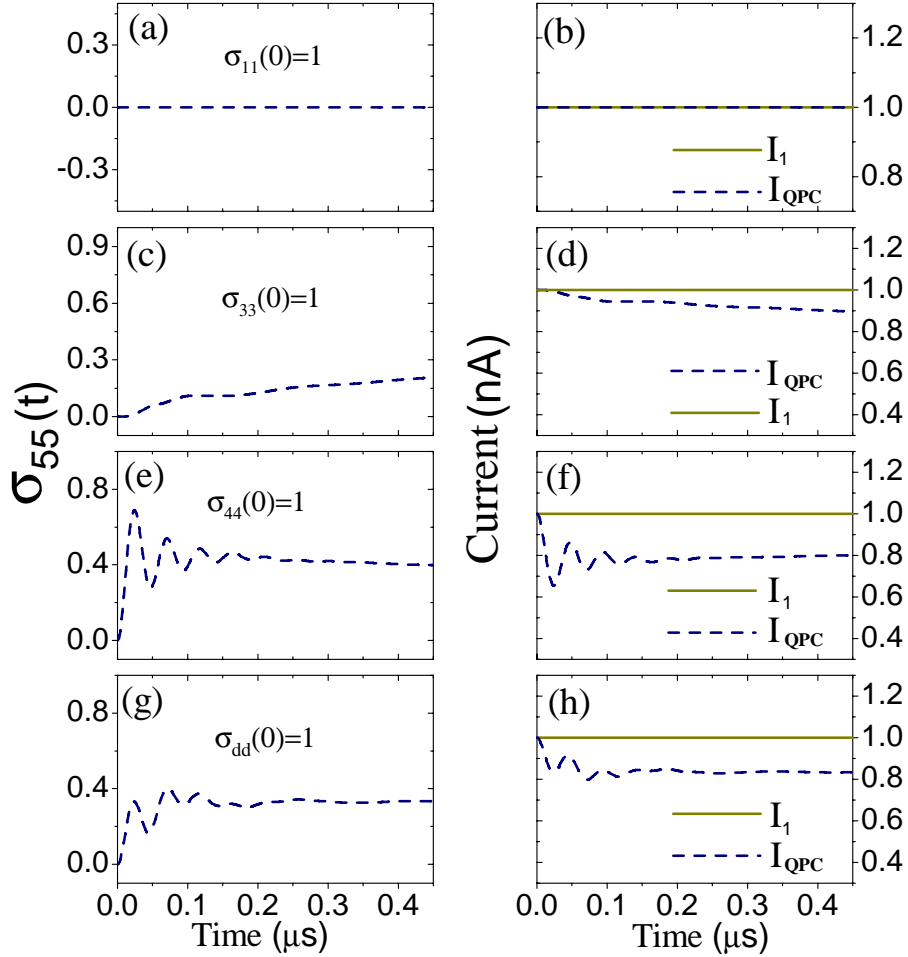}}
\caption{(Color online). Time evolution of the occupation
probability $\sigma_{55}(t)$ of the single-dot singlet state
$|S_S\rangle$ for four different initial conditions: (a)
$\sigma_{11}(0)=1$, (c) $\sigma_{33}(0)=1$, (e) $\sigma_{44}(0)=1$
and (g) $\sigma_{dd}(0)=1$, where the hyperfine interaction is
included. (b), (d), (f) and (h) The corresponding QPC currents for
these four different initial conditions. The nuclear magnetic fields
are chosen to be $B_{\rm{NL}}^{x,y,z}=(-2,1, 3)$~m$\textbf{T}$, and
$B_{\rm{NR}}^{x,y,z}=(-1, 2, 0)$~m$\textbf{T}$. The other parameters
are the same as in Figure.~\ref{fig5}.}\label{fig6}
\end{figure}

We first consider the case without hyperfine interactions, i.e.,
$\mathbf{B}_{\rm{NL(R)}}=0$. In this limit, the coupling between the
states $|3\rangle$ and $|4\rangle$ vanishes. (i) If initially the
DQD takes the double-dot triplet state $|1\rangle$, the system will
not evolve into other states due to the large Zeeman splitting. In
this case, the electron in the left dot does not hop into the right
one and the current through the QPC does not change, as shown in
Figures.~\ref{fig5}(a) and \ref{fig5}(b). (ii) Alternatively, for an
initial state $|3\rangle$, the DQD system will also remain at this
state because $|3\rangle$ does not couple with any other states.
Similar to the case of the initial state $|1\rangle$, the occupation
probability of the single-dot singlet state $|5\rangle$ is zero and
the current through the QPC also remains unchanged [see
Figures.~\ref{fig5}(c) and \ref{fig5}(d)]. (iii) In contrast, as
shown in Figures.~\ref{fig5}(e) and \ref{fig5}(f), if the DQD system
initially stays at $|4\rangle$, it couples with the single-dot
singlet state due to the hopping coupling between the two dots. This
gives rise to nonzero occupation probability for the single-dot
singlet state and a variable current through the QPC. (iv) The
results shown in Figures.~\ref{fig5}(g) and \ref{fig5}(h) look like
a combination of the results in both (ii) and (iii). This is because
the initial state $|d\rangle=|\!\downarrow_L\uparrow_R\rangle$ is a
superposition of the states $|3\rangle$ and $|4\rangle$
[cf.~Eq.~(\ref{superposition})]. Moreover, only the state
$|4\rangle$ contributes to the variations of both the probability of
the state $|5\rangle$ and the current through the QPC.

Moreover, it is shown in Figures~\ref{fig5}(b) and \ref{fig5}(d)
that the two cases with initial states $|1\rangle$ and $|3\rangle$
are indistinguishable in measuring the electron spin. This is due to
neglecting the hyperfine interactions. When they are included, these
two cases become distinguishable (cf.~Figure.~\ref{fig6}).

For an initial state $|1\rangle$ or $|4\rangle$, the results are
similar to those without the hyperfine interactions. This can be
clearly seen by comparing Figures.~\ref{fig6}(a) and \ref{fig6}(b)
with Figures.~\ref{fig5}(a) and \ref{fig5}(b) for initial state
$|1\rangle$, and similarly comparing Figures.~\ref{fig6}(e) and
\ref{fig6}(f) with Figures.~\ref{fig5}(e) and \ref{fig5}(f) for
initial state $|4\rangle$. In contrast, for initial state
$|3\rangle$, because $|3\rangle$ and $|4\rangle$ are degenerate,
hyperfine interactions are able to provide significant couplings.
Moreover, state $|4\rangle$ is also coupled to $|5\rangle$ via
hopping. Thus, from initial state $|3\rangle$, the system can
finally evolve to $|5\rangle$. Indeed, this is reflected in the
variations of both the occupation probability of state $|5\rangle$
and the QPC current [Comparing Figures.~\ref{fig6}(c) and
\ref{fig6}(d) with Figures.~\ref{fig5}(c) and \ref{fig5}(d)].
For the initial state $|d\rangle=|\!\downarrow_L\uparrow_R\rangle$,
the probability of the state $|5\rangle$ and the QPC current [shown
in Figures \ref{fig6}(g) and \ref{fig6}(h)] also look like a
combination of the results for both the initial states $|3\rangle$
and $|4\rangle$ [shown in Figures~\ref{fig6}(c)-\ref{fig6}(f)],
similar to the case without hyperfine interaction in the DQD.

\begin{figure}
\epsfxsize 12cm \centerline{\epsffile{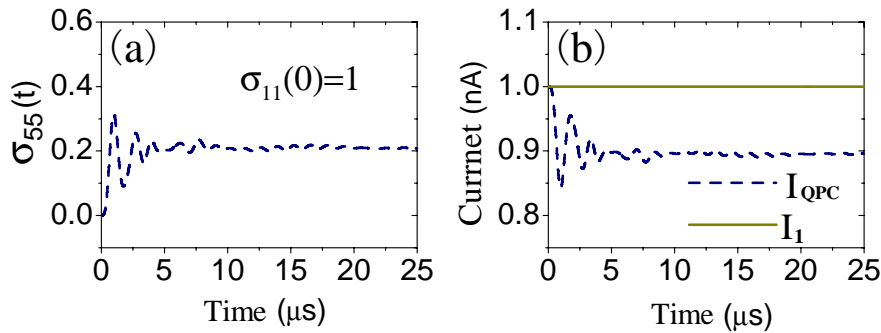}}
\caption{(Color online). (a) Time evolution of the occupation
probability of the single-dot singlet state in the presence of an
ESR magnetic field for the initial state $|1\rangle$ (see the text).
(b) The corresponding QPC current.}\label{fig7}
\end{figure}

The last issue to be addressed is to determine if the electron is
successfully injected into the left dot. If the injection fails, the
QPC current is always $I_1$. The result is the same as that with a
spin-up electron injected into the qubit dot. To distinguish between
these two cases, we can apply a transverse magnetic field to flip
the electron spin in the qubit dot. For a successful injection, the
electron with spin up will flip to become spin down and then hop
into the right dot. The DQD system then takes the state
$|\downarrow_L\uparrow_R\rangle$. As discussed above, this state,
i.e., the superposition state of $|3\rangle$ and $|4\rangle$, is
coupled to the single-dot singlet state $|5\rangle$, giving rise to
a variation of the occupation probability $\sigma_{55}$ [see
Figure.~\ref{fig7}(a)] as well as the QPC current [see
Figure.~\ref{fig7}(b)]. This is different from the case of a
constant current in the absence of any successful electron injection
into the left dot.

\section{Conclusion}

In summary, we have studied the readout of a single electron spin in
a DQD system. The electron spin is initially confined in the QD
serving as a qubit dot. A reference dot is coupled to the qubit dot
via a tunneling coupling. Also, a QPC acts as a measurement device,
placed near the reference dot for detecting the variation of the
electron number in the reference dot. We have considered the two
implementations in which either one or two electrons occupy the DQD.
In the one-electron case, the only electron in the DQD is the qubit
electron to be measured. An external magnetic field is applied to
both dots so that the energy level splittings $\varepsilon_\uparrow$
and $\varepsilon_\downarrow$ for spin-up and spin-down electrons are
different. Gate voltages of the two dots are tuned so that
$\varepsilon_\uparrow\sim 0$ and
$\varepsilon_\downarrow\gg\Omega_0$. These conditions ensure that
only a spin-up electron but not a spin-down electron in the qubit
dot can tunnel to the reference dot. This gives rise to very
different currents through the QPC and can be used to readout the
electron spin states of the qubit dot. In the two-electron case, an
additional spin-up electron is always confined in the reference dot.
This can be easily achieved by properly tuning the gate voltages of
the dots. We have shown that the electron spin states of the qubit
dot can also be readout by applying an external magnetic field when
considering effects of hyperfine interactions between the electron
spin and the nuclear spins of the host materials. In the high Zeeman
splitting limit, the flipping of the electron spin induced by the
hyperfine interactions are greatly suppressed. In this case, only a
spin-down electron in the qubit dot can tunnel to the reference dot.
This again allows one to distinguish between the electron spin
states in the qubit dot by measuring the currents through the QPC.
Furthermore, we propose an approach involving an ESR oscillating
magnetic field which can confirm the success of an electron
injection event into the qubit dot.

\section*{Acknowledgments}
This work was supported by the SRFDP, the PCSIRT, the NFRPC grant
No. 2006CB921205 and the National Natural Science Foundation of
China grant Nos. 10534060 and 10625416.

\end{document}